# 'Digital' Electron Diffraction - Seeing the Whole Picture

Richard Beanland,[a] Paul J Thomas[b] David I Woodward,[a] Pamela A Thomas[a] and Rudolf A Roemer[a]

[a]Department of Physics, University of Warwick, Coventry CV4 7AL, UK, and [b]Gatan UK Ltd, 25 Nuffield Way, Abingdon, Oxon, OX14 1RL, UK.

E-mail: r.beanland@warwick.ac.uk

**Synopsis**  Computer control of beam tilt and image capture allows the collection of electron diffraction patterns over a large angular range, without any overlap in diffraction data and from a region limited only by the size of the electron beam. This results in a significant improvement in data volumes and ease of interpretation.

**Abstract**  The advantages of convergent beam electron diffraction for symmetry determination at the scale of a few nm are well known. In practice, the approach is often limited due to the restriction on the angular range of the electron beam imposed by the small Bragg angle for high energy electron diffraction, i.e. a large convergence angle of the incident beam results in overlapping information in the diffraction pattern. Techniques have been generally available since the 1980s which overcome this restriction for individual diffracted beams, by making a compromise between illuminated area and beam convergence. Here, we describe a simple technique which overcomes all of these problems using computer control, giving electron diffraction data over a large angular range for many diffracted beams from the volume given by a focused electron beam (typically a few nm or less). The increase in the amount of information significantly improves ease of interpretation and widens the applicability of the technique, particularly for thin materials or those with larger lattice parameters.

## 1. Introduction

The weak interaction of X-rays and neutrons with matter makes them ideal for structure solution of bulk materials (of size $\gtrsim 10\mu m$) since single scattering events dominate, but results in a low scattering intensity from small volumes. Conversely, the strong interaction of electrons with matter allows analysis of nanoscale volumes, but complicates their use due to the dominance of multiple (dynamical) scattering events. The specimen must be very thin (typically < 200 nm) to allow transmission of the electron beam, usually in a transmission electron microscope (TEM). Dynamical scattering causes the diffracted intensity for any given reflection *hkl* from a crystalline material to vary enormously as a function of the incident beam orientation, even when the Bragg condition is satisfied exactly. It also produces significant intensity in reflections that are completely absent in singly scattered diffraction (i.e. 'forbidden' reflections). This broadly prevents the simple use of electron diffraction patterns for structure solution. Nevertheless, the symmetry of an electron diffraction pattern is still determined by the symmetry of the crystal from which it is produced, and dynamical scattering has some distinct advantages, such as information describing the phase of the diffracted electrons, (Spence, 1993, Tanaka & Tsuda, 2011) sensitivity to chirality (Johnson, 2007) and the breaking of Friedel's law (Friedel, 1913, Steeds & Vincent, 1983) which render X-ray and neutron scattering insensitive to the presence of a centre of symmetry in a crystal. These factors, together with a greater sensitivity to valence electron densities, (Spence, 1993, Zuo, 2004) mean that electron diffraction data is in principle richer and more sensitive than that from other techniques.

The description of electron diffraction using dynamical scattering theory is well established, and the difficulties do not lie in a lack of a well-understood theory or low signal strength; rather, the main challenge is often to extract a sufficient *quantity* of data to allow dynamical theory to be applied with confidence. At the heart of the problem is the fact that, because of the very small wavelength of high-energy electrons, Bragg angles, $\theta_{hkl}$, for diffracted electron beams are small (typically less than 1°), while diffraction can occur for most strong reflections at large deviations (>2 $\theta_{hkl}$ or more) from the Bragg condition. This inevitably leads to overlapping diffracted beams unless the half-convergence angle, $\alpha$, of incident illumination is restricted to be less than the smallest Bragg angle in any given convergent beam electron diffraction (CBED) pattern, a fact which has been appreciated since the very beginning of electron diffraction. (Kossel & Möllenstedt, 1939) This "overlap problem" severely restricts the angular range of data that can be obtained, particularly from materials with relatively large lattice parameters.

This problem was partially solved by Tanaka (Tanaka *et al.*, 1980) using a highly convergent beam and a displacement of the specimen from the image plane of the objective lens in a TEM, blocking all diffracted beams apart from the one of interest by placing an aperture in a conjugate image plane. A slightly different solution was developed by Eades (Eades, 1980) using a parallel beam rocking both above and below the specimen in scanning (STEM) mode. Both of these approaches give access to a more complete diffraction dataset for one diffracted beam. However, such large angle convergent beam (LACBED) patterns obtained using the Tanaka method can only be obtained from large, parallel-sided, flat crystals (Tanaka & Tsuda, 2011) (any bending of the crystal leads to distortions in the pattern) and the scanning

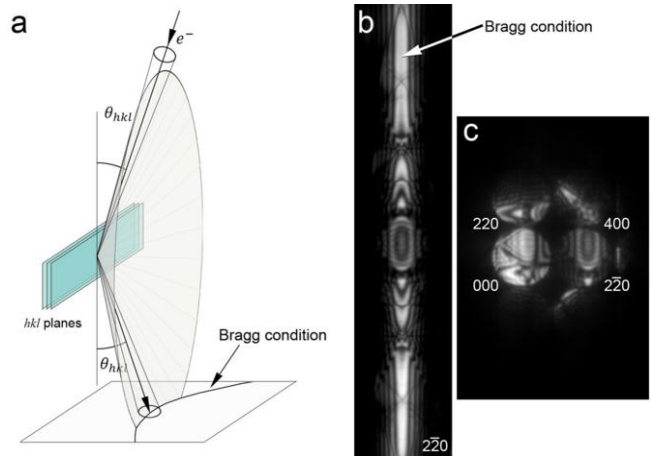

**Figure 1** (a) The geometry of large-angle convergent beam electron diffraction (LACBED) for one diffracted beam, ignoring all other diffracted beams. The Bragg condition is satisfied when the incident and exit beams make an angle $\theta_{hkl}$ to the diffracted planes, defining a cone. This gives a parabola on the diffraction pattern which, because of the very small Bragg angle, appears as a straight line in a LACBED pattern (b), taken from a [100]silicon crystal. (c) the corresponding CBED pattern, with a small part of the LACBED pattern visible in the $2\bar{2}0$ disc method is difficult to implement. Furthermore, both of these techniques only allow access to one diffracted beam at a time; obtaining several LACBED images is both time-consuming and requires significant effort and skill on the part of the operator. Recently, Koch (Koch, 2011) used computer control of the microscope to rock a parallel beam in a similar way to Eades, together with partial compensation of the tilt below the specimen to produce, a large number of low-resolution LACBED patterns, again captured in a single exposure on camera. Thus to date, almost all[1] electron diffraction techniques sample only a very limited part of the full dynamical diffraction dataset.

Here, we describe a method which eliminates the fundamental problem of beam overlap using computer control, providing large diffraction datasets with many diffracted beams which contain detailed information from a region as small as the electron beam focused on the specimen, which can be a few nm in size or less.

## 2. Methods

If the problem of beam overlap can be neglected, a dark-field LACBED pattern takes the form of a bright line of diffracted intensity, corresponding to the angle at which the Bragg condition is satisfied (Fig. 1). A small portion of this pattern is visible in one disc of a CBED pattern, Fig. 1(c). It is immediately apparent that by collecting a large number of individual CBED patterns with different incident beam tilts, the LACBED pattern can be reconstructed

---

[1] Access to diffraction patterns which contain *both* a large angular range *and* multiple diffracted beams was demonstrated by Terauchi, M. & Tanaka, M. (1985). *Journal of Electron Microscopy* **34**, 128-135. using a TEM in an unusual configuration. This feat does not appear to have been repeated since.

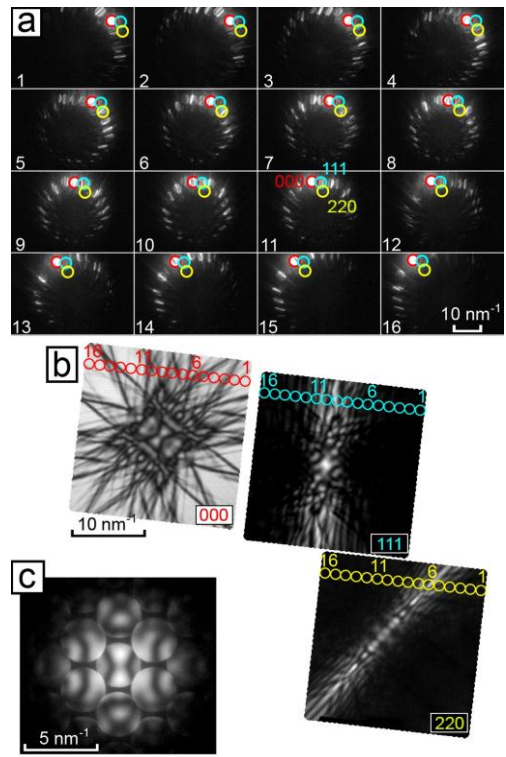

**Figure 2** (a) Sixteen CBED patterns from [110] silicon with varying beam tilts. The 000 (red), 111 (blue) and 220 (yellow) beams are highlighted in each. (b) Digital reconstruction of LACBED patterns from many individual CBED patterns, highlighting the components from the patterns in (a). (c) The on-axis CBED pattern.

by combining the relevant parts of each individual CBED pattern as shown in Fig. 2. We implemented this approach using a JEOL2100 TEM with standard computer control of the electron-optic lenses and a digital camera. Standard conditions for CBED were used, i.e. the electron beam was focused to a small probe (typically ~15nm FWHM) on a thin specimen, with the illumination convergence angle adjusted such that there was no significant overlap of the discs in the diffraction pattern. The tilt of the incident beam was controlled via a computer script to scan over a large angular range (typically up to 0.1 radians, or ~5.7°, corresponding to approximately 50 nm$^{-1}$, and a diffraction pattern was collected at each different incident beam tilt using the CCD camera. The beam tilt step was adjusted to give ~30% overlap between consecutive patterns. The exposure time for an individual CBED pattern was typically 40ms, although camera processing overhead increased the time between individual frames to approx. 80ms (i.e. 1000 patterns in 80 seconds - sufficiently fast to avoid problems with specimen drift or contamination). In this microscope the upper limit of the tilt range which can be obtained without computer control is due to the spherical aberration of the pre-field objective lens, causing shifts and changes in beam shape for incident beam tilts much more than 60-100 mrad from the optic axis, depending upon the excitation of the final condenser mini-lens. Careful alignment of beam tilt compensation was employed to ensure that beam shift during data acquisition was less than 20% of the FWHM of the electron beam (~3 nm) for the maximum beam tilt used in any given dataset. The data from each different diffracted beam were then recombined into a single image using a second computer script, giving a montage of D-LACBED patterns. For this angular range, useful D-LACBED patterns can be extracted for typically 50-60 different reflections from a single dataset.

## 3. Results

We begin with data from 'standard' materials GaAs and Si, which have often been used for conventional CBED investigations. Figure 3 shows the central seventeen patterns obtained from [1$\bar{1}$0] GaAs. The D-LACBED patterns are arranged such that they have the same relative positions as in the conventional electron diffraction patterns, although note that each D-LACBED pattern covers an angular range similar to the whole of the conventional patterns shown for comparison to the left. The relationship between the symmetries of an electron diffraction pattern and that of the crystal was determined by Buxton et al., (Buxton *et al.*, 1976) and is based upon the premise that all of the information visible in Fig. 3 is available. Normally, when performing such a symmetry determination using CBED, the skill and time needed carefully to choose the specimen thickness, as well as manual tilting of the incident beam and/or specimen to allow different parts of each dark field pattern to be observed, is considerable. The ease of a single click for data collection, and the significant improvement

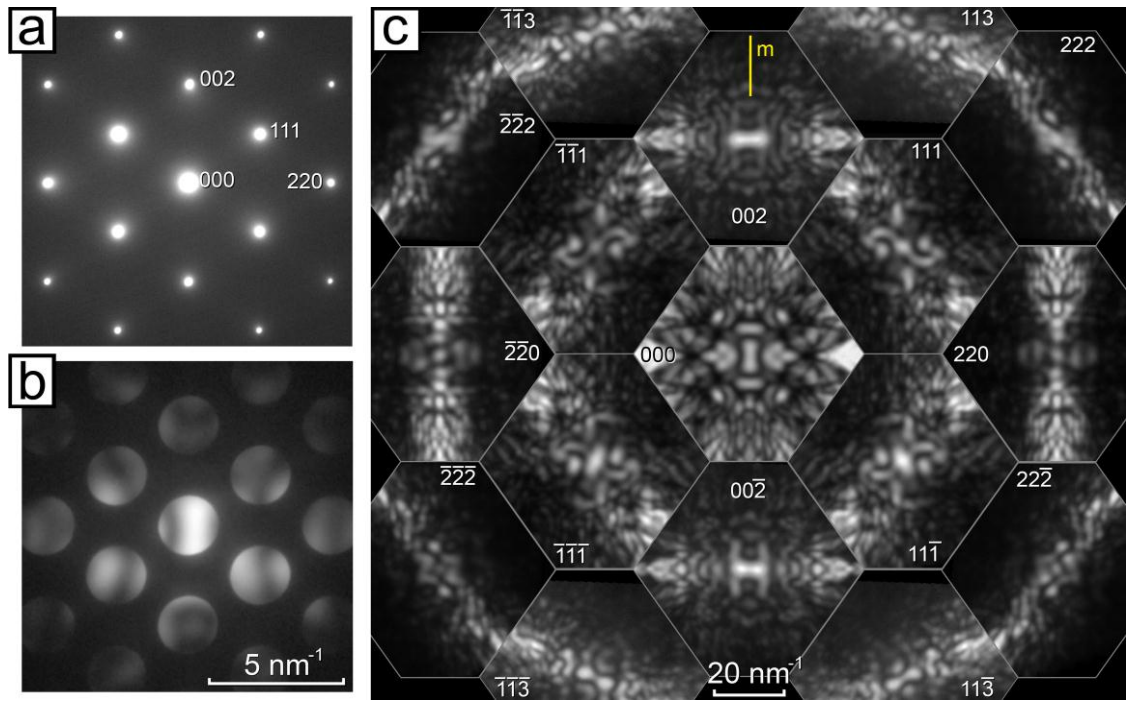

**Figure 3** (a) SAED, (b) CBED and (c) montage of 17 D-LACBED patterns taken from [1$\bar{1}$0] GaAs; the patterns are arranged in positions corresponding to the CBED pattern. The diffraction vector **g** is indicated for each pattern and the (110) mirror plane is indicated by the letter *m*; no horizontal mirror is present. The whole pattern symmetry is $m1_R$.

in symmetry identification that results from access to a larger part of the complete dynamical diffraction dataset, is readily apparent.

The directly transmitted beam may in general have higher symmetry than the pattern as a whole, and this is the case here with the **g** = 000 D-LACBED pattern having symmetry 2*mm*. Each individual dark field pattern, corresponding to a different diffracted beam, can have symmetry up to 2*mm* in itself; in Fig. 2 it can be seen that this is only the case for those patterns crossing the vertical (110) mirror plane, i.e. **g** = 002-type patterns. All other patterns, i.e. **g** = 111, 222, 220 and 113-type, have 2-fold symmetry about their centre. This symmetry operation, denoted $1_R$ (Buxton et al., 1976), can indicate the presence of a mirror plane perpendicular to the electron beam; however it is also present in all zero-order Laue zone reflections (as is the case here) since the projected potential of the crystal is independent of the sense of the electron beam direction. There is no horizontal (001) mirror present, indicating the polarity of the crystal, and the lack of equivalence in ±**g** pairs, indicating the lack of a centre of symmetry, is obvious. As a whole, the pattern has $m1_R$ symmetry, as expected for a crystal with space group $F\bar{4}3m$ and point group $\bar{4}3m$.

The multiple scattering processes which are inherent to electron diffraction, in combination with a limited sampling of the dynamical diffraction dataset, often give rise to the impression that electron diffraction is unreliable or limited in application in comparison with X-ray crystallography. However when the structure is well known, as in the case of GaAs, it is straightforward to reproduce the experimental data using standard simulation

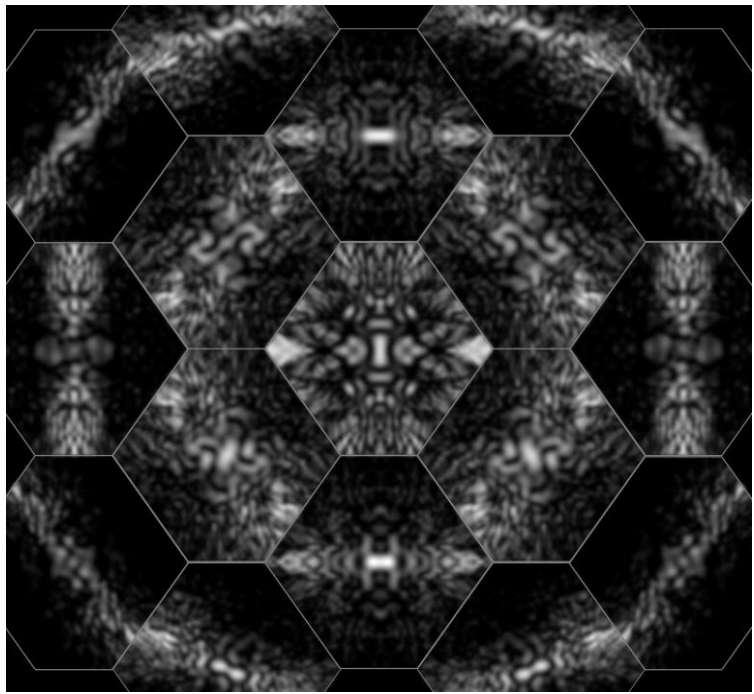

**Figure 4**  Montage of simulated LACBED patterns corresponding to the experimental data of Fig. 3 at a specimen thickness of 65 nm.

software (Stadelmann, 1987), as illustrated by Fig. 4.

A similar D-LACBED montage taken from [1$\bar{1}$0] silicon, with space group F$d\bar{3}m$ and point group $m\bar{3}m$ is shown in Figure 4. The equivalence between the two fcc sub-lattices in the diamond structure doubles the number of symmetry elements in the crystal space group (including the addition of a centre of symmetry, giving an equivalence between ±**g** pairs) but also leads to forbidden reflections with indices **g** = 002, 222, 442... . In conventional electron diffraction these forbidden reflections often appear just as strongly as the 'allowed' reflections, as can be seen in Fig. 3a. Nevertheless, the 002 reflection should drop to zero intensity at angles sufficiently far away from a zone axis (as employed in precession electron diffraction (Vincent & Midgley, 1994)) where multiple scattering pathways do not exist. This is indeed the case and is clearly visible in D-LACBED data, as shown in Fig. 3c. In general we find that forbidden reflections are readily identified in D-LACBED datasets, even when the crystal is relatively thick and multiple scattering dominates.

While it is useful to see the large improvement in data available when examining GaAs and Si, the real utility of the technique lies in its application to nanostructured materials which are difficult to tackle using X-ray diffraction, or even conventional electron diffraction. We therefore consider a material which has unknown symmetry; NaBiCaTeO$_6$, an ($A^{3+}A^{1+}$)$B^{2+}$TeO$_6$ material that we take here to be an example of a typical perovskite oxide. The prototype perovskite structure is cubic, with symmetry $Pm\bar{3}m$ and lattice parameter typically around 0.4nm; NaBiCaTeO$_6$ might be expected to exhibit *A* and/or *B* cation ordering, and/or displacements from nominal positions in the unit cell, and/or tilting of the oxygen octahedra, (Glazer, 1972) or any combination of these effects (Howard & Stokes, 2004, Howard & Zhang, 2004, Kishida *et al.*, 2009). In any case, we expect the space group to be some sub-group of $Pm\bar{3}m$. Tellurium-containing compounds can exhibit ferroelectric or antiferroelectric behaviour (Venevtsev *et al.*, 1974, Politova & Venevtsev, 1975); in terms of functional properties, ferroelectric behaviour is preferable since this leads to piezoelectric, pyroelectric and other useful applications. Since these ferroic properties only exist in materials without a centre of symmetry, determination of the crystal point and space group has direct relevance to technological utility, and electron diffraction has a distinct advantage here. *A-priori* determination of crystal space group from dynamical electron diffraction patterns has been described by Goodman (Goodman, 1975), Steeds and Vincent, (Steeds & Vincent, 1983), Tanaka (Tanaka & Tsuda, 2011), and more recently (Morniroli *et al.*, 2012, Jacob *et al.*, 2012)all of whom rely on the original classification of dynamic diffraction symmetries of Buxton et al.(Buxton *et al.*, 1976). As the structure is unknown, we will use a pseudo-cubic notation (i.e. treat the crystal as if it were a prototype perovskite, for indexing purposes only). Data were collected from defect-free domains in a polycrystalline ceramic, prepared for TEM using standard techniques and similar probe sizes were used as in the Si and GaAs examples. However, significantly smaller convergence angles were required due to the larger lattice parameter encountered.

Figure 5 shows electron diffraction patterns from $[001]_{PC}$. Half-order ½ $h$ $2k$ $2l$ ('superstructure') spots are visible in the SAED pattern, where $h$, $k$, $l$ are odd integers, often described as half 'odd-even-even' or ½ *oee* spots.(Woodward & Reaney, 2005, Reaney *et al.*, 1994), indicating a doubling of periodicity along $[100]_{PC}$ but not $[010]_{PC}$. In the CBED discs of Fig. 5b, some bright and dark regions are present with no apparent symmetry, but it is not clear if this is simply because the crystal is not aligned exactly with the incident electron beam. The information available is rather limited. Conversely, much more information is easily extracted from the montage of Fig. 5c, which clearly shows a lack of any mirror symmetry, two-fold symmetry in all individual D-LACBED patterns (due to projection) and $2 1_R$ in the pattern as a whole. It is clear that (sub) unit cell distortions and/or cation ordering have broken all $\{110\}_{PC}$, and $\{100\}_{PC}$ mirrors that could be present in this pattern. The $2_R$ operation is consistent with the presence of a 2-fold along the beam direction, or a centre of symmetry in the crystal, or both; the lack of any 3D information in the form of HOLZ lines prevents them from being distinguished in this case.

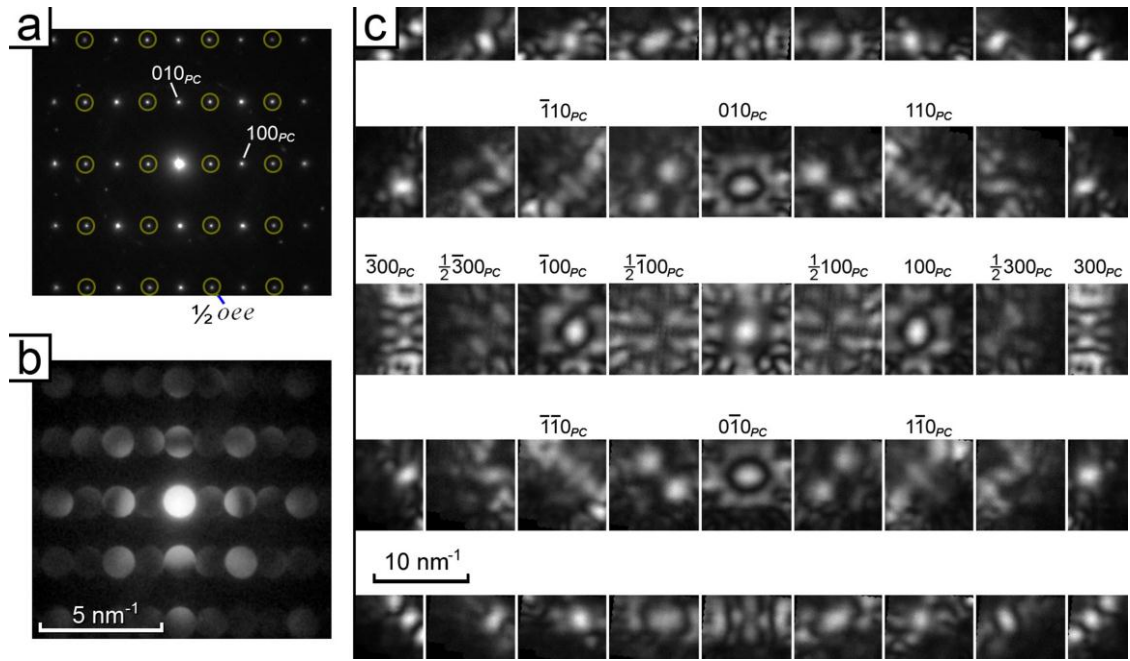

**Figure 6** (a) SAED, (b) CBED and (c) montage of 45 D-LACBED patterns taken from $[001]_{PC}$ NaBiCaTeO$_6$. The pattern symmetry is $21_R$.
Highlighted reflections in the SAED pattern are forbidden and would have zero intensity without multiple scattering. (c) shows the 002 D-LACBED pattern at a large angle from the zone axis and it is clear that the intensity does indeed drop to zero. The whole pattern symmetry is $2mm1_R$.

Figure 6 shows a similar trio of diffraction patterns from a $[111]_{PC}$ axis. Here, ½ *ooe* superstructure spots are visible in the SAED pattern; again there is little detail in the CBED discs, and the superstructure discs are weak but visible. Once more, an enormous amount of information is visible in the D-LACBED pattern. Strikingly obvious black crosses are visible in alternate patterns along the horizontal systematic row, i.e. ½ $0\bar{1}1_{PC}$ and ½ $0\bar{3}3_{PC}$ type patterns. They are also present along the vertical ½ $\overline{112}_{PC}$ systematic row. These dark crosses are dynamical extinction effects, (Gjønnes & Moodie, 1965, Tanaka *et al.*, 1987) often known as Gjønnes-Moodie lines, and are present when the incident electron beam is parallel to a glide plane or perpendicular to a $2_1$ screw axis. Whilst in principle these could be observed in CBED patterns, this is difficult when the crystal is thin and the beam convergence angle is small; not enough of the diffraction dataset is visible. It is interesting to compare these patterns with the ½ $100_{PC}$ D-LACBED patterns in Fig. 5, which also contain a dark cross – however the ½ $300_{PC}$ D-LACBED patterns have no black cross and we therefore do not consider them to indicate the presence of a glide plane or screw axis.

The D-LACBED data from these two zone axes – and the knowledge that the crystal is a perovskite – are sufficient to determine the point group of the crystal. The $2_1$ element along $[\bar{1}10]_{PC}$ and perpendicular glide plane give a minimal point group of $2/m$. However the lack of any symmetry elements at the $[\bar{1}10]_{PC}$ zone axis (apart from the possible presence of a 2-

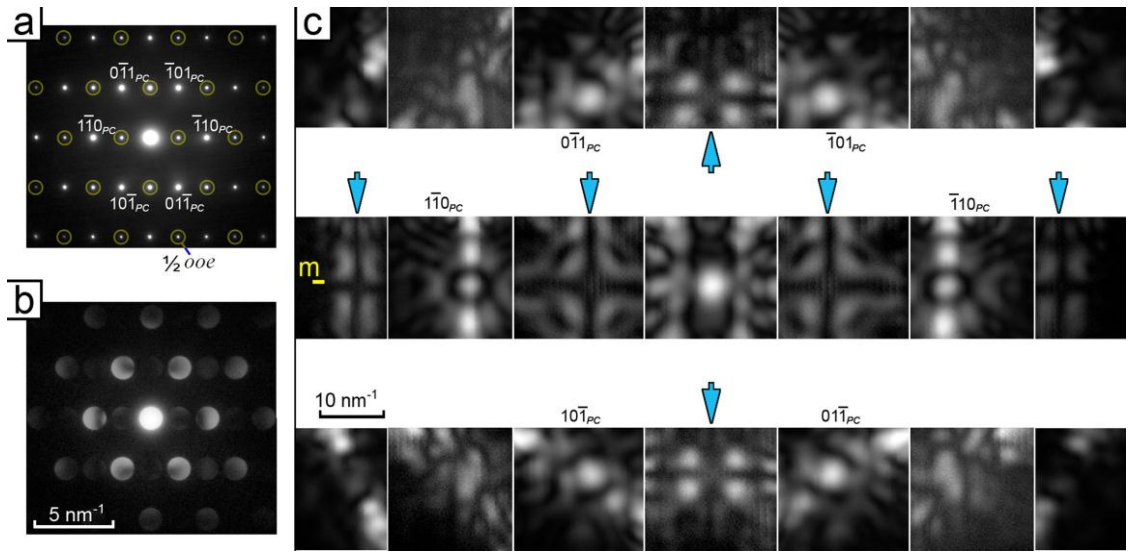

**Figure 7** (a) SAED, (b) CBED and (c) montage of 21 D-LACBED patterns taken from $[111]_{PC}$ NaBiCaTeO$_6$. Arrows mark Gjønnes-Moodie lines, indicating the presence of a $2_1$ axis along $[\bar{1}10]_{PC}$ and a $(\bar{1}10)_{PC}$ mirror-glide plane. The pattern symmetry is $2mm1_R$.

fold axis or centre of symmetry) rules out any larger point group; there is no point group which has a 2-fold axes along [100] and $[\bar{1}10]$, a $(\bar{1}10)$ mirror without further symmetry elements which would affect the [100] pattern. The only possibility is that the $2_{1R}$ D-LACBED pattern symmetry of Fig. 5 is due to the centre of symmetry in the point group $2/m$. A third pattern is required to determine the translation vector of the glide plane, since dark Gjønnes-Moodie crosses are expected in the zero-order Laue zone if there is any component of the glide vector perpendicular to the electron beam. We thus tilted the crystal from the $[111]_{PC}$ orientation about the $2_1$ axis to the $[110]_{PC}$ zone axis (Figure 7). The Gjønnes-Moodie crosses remain (as expected) along the $2_1$ axis $[\bar{1}10]_{PC}$, and are also present along the perpendicular direction $[001]_{PC}$; the glide translation thus cannot be parallel to the $[110]_{PC}$ zone axis and can therefore only be parallel to $[001]_{PC}$. In SAED patterns, superstructure spots indicate a doubling of the *P* lattice along all pseudo-cubic axes. This fixes the space group as #14, *P*2$_1$/*c*, with the unique *b*-axis parallel to $[\bar{1}10]_{PC}$ and the *c*-axis parallel to $[001]_{PC}$. The ease of determining space group in this example is a direct result of the greater level of detail available in D-LACBED patterns in comparison with other diffraction techniques.

## 4. Discussion

We have demonstrated that computer control of beam tilt and image capture in a TEM can be used to overcome the problem of overlapping diffracted beams, quickly providing very rich diffraction datasets which can be used for easy determination of crystal symmetry on a nanometre scale. This approach stems from an appreciation of the fact that an image gathered

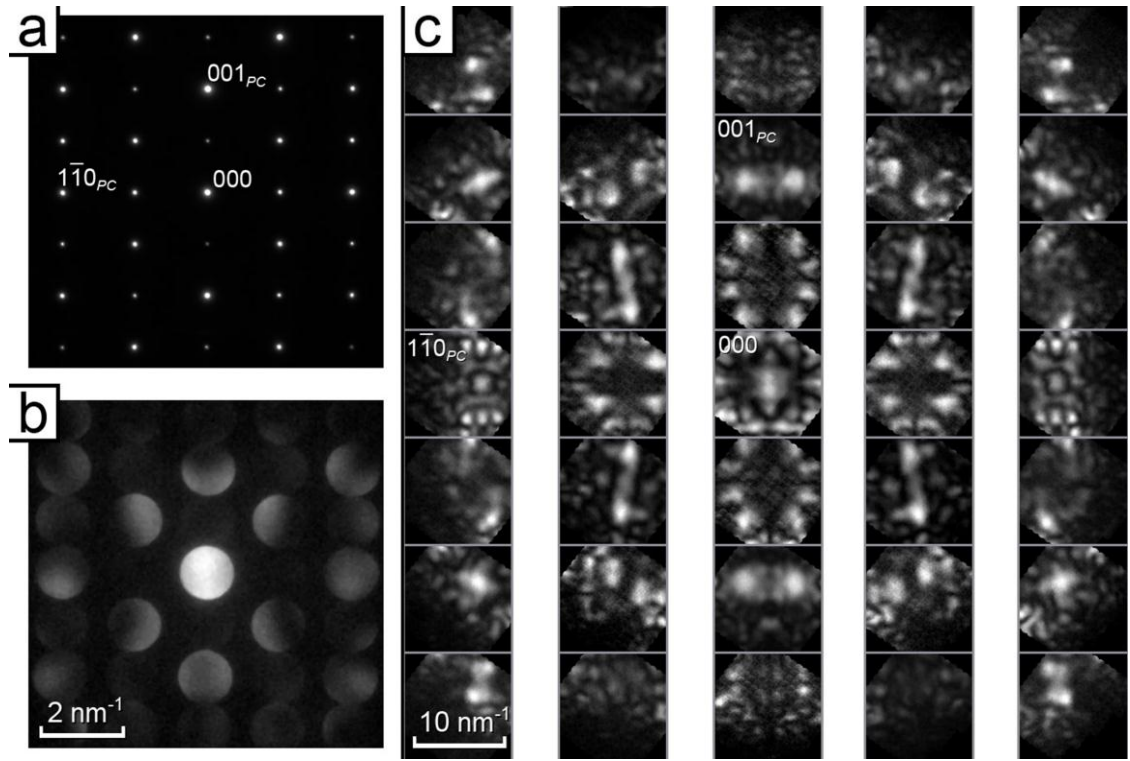

**Figure 8** (a) SAED, (b) CBED and (c) montage of 35 D-LACBED patterns taken from $[110]_{PC}$ NaBiCaTeO$_6$. Gjønnes-Moodie dark crosses again mark the presence of a $2_1$ axis along $[\bar{1}10]_{PC}$ and a $(\bar{1}10)_{PC}$ mirror-glide plane.

from a CCD camera is a numerical dataset which is easily combined with other datasets. The greatest experimental difficulty is to gather a suitable amount of data in a reasonable time, since specimen drift and contamination would render the dataset meaningless. This is achieved using low-level programming to optimize the capture rate of CBED patterns. We typically achieve capture rates >10 patterns per second, most of the D-LACBED patterns shown here being a combination of up to a thousand individual CBED patterns, acquired in less than 120 seconds. It is clear that optimization of image capture and microscope control could easily improve on this, potentially reducing data collection times by an order of magnitude or more (Humphry *et al.*, 2012).

It is our hope that this technique will become the tool of choice for investigation of local symmetry and structure using electron diffraction, supplementing standard CBED techniques and finding a host of applications across many materials systems. The understanding gained of dynamical electron diffraction patterns (Gjonnes & Moodie, 1965, Buxton *et al.*, 1976, Goodman, 1975, Tanaka & Tsuda, 2011, Morniroli *et al.*, 2012) still applies to these new diffraction datasets, and the significant extra detail in D-LACBED patterns allows immediate and unambiguous determination of the presence of symmetry elements. Here, we have deliberately chosen a 'standard' TEM without energy filtering or spectroscopy, and without even the smaller and more intense probe afforded by a field emission electron gun. There is no fundamental barrier to implementation of this technique on higher performance machines; the sub-nm probe available on aberration-corrected machines should allow investigation of local symmetries close to the unit cell level, although close control of beam shape, size and position will be necessary (Koch, 2011). Furthermore, the use of energy filtering, while unnecessary for the symmetry determination described here, produces more quantitative data. Looking forward, the increase in data quantity and quality produced by D-LACBED may also allow quantitative analysis of diffracted intensities to determine valence electron distributions (Zuo, 2004) to be performed on a wider range of materials, opening up the exciting possibility of examining strongly-correlated systems (for example, high-$T_c$ superconductors). Furthermore, we note that all other types of electron diffraction, such as SAED, CBED and even precession electron diffraction (PED) patterns, are a smaller sample of the more complete D-LACBED dataset, and can be derived in a simple and straightforward manner from the 'digital' electron diffraction patterns shown here.

**Acknowledgements** This work was funded by the EPSRC under grant number EP/J009229/1

**References**

Buxton, B. F., Eades, J. A., Steeds, J. W. & Rackham, G. M. (1976). *Philosophical Transactions of the Royal Society of London. Series A, Mathematical and Physical Sciences* **281**, 171-194.
Eades, J. A. (1980). *Ultramicroscopy* **5**, 71-74.
Friedel, G. (1913). *C.R. Acad. Sci. Paris* **157**, 1533-1536.
Gjonnes, J. & Moodie, A. F. (1965). *Acta Crystallographica* **19**, 65-67.
Glazer, A. (1972). *Acta Crystallographica Section B* **28**, 3384-3392.
Goodman, P. (1975). *Acta Crystallographica Section A* **31**, 804-810.
Howard, C. J. & Stokes, H. T. (2004). *Acta Crystallographica Section B* **60**, 674-684.
Howard, C. J. & Zhang, Z. (2004). *Acta Crystallographica Section B* **60**, 249-251.
Humphry, M. J., Kraus, B., Hurst, A. C., Maiden, A. M. & Rodenburg, J. M. (2012). *Nat Commun* **3**, 730.
Jacob, D., Ji, G. & Morniroli, J. P. (2012). *Ultramicroscopy* **121**, 61-71.
Johnson, A. (2007). *Acta Crystallographica Section B* **63**, 511-520.
Kishida, K., Goto, K. & Inui, H. (2009). *Acta Crystallographica Section B* **65**, 405-415.
Koch, C. T. (2011). *Ultramicroscopy* **111**, 828-840.
Kossel, W. & Möllenstedt, G. (1939). *Annalen der Physik* **428**, 113-140.
Morniroli, J. P., Ji, G. & Jacob, D. (2012). *Ultramicroscopy* **121**, 42-60.
Politova, E. D. & Venevtsev, Y. N. (1975). *Mater. Res. Bull.* **10**, 319-325.
Reaney, I. M., Colla, E. L. & Setter, N. (1994). *Jpn. J. Appl. Phys. Part 1 - Regul. Pap. Short Notes Rev. Pap.* **33**, 3984-3990.
Spence, J. (1993). *Acta Crystallographica Section A* **49**, 231-260.
Stadelmann, P. A. (1987). *Ultramicroscopy* **21**, 131-145.
Steeds, J. W. & Vincent, R. (1983). *Journal of Applied Crystallography* **16**, 317-324.
Tanaka, M., Saito, R., Ueno, K. & Harada, Y. (1980). *Journal of Electron Microscopy* **29**, 408-412.
Tanaka, M., Terauchi, M. & Sekii, H. (1987). *Ultramicroscopy* **21**, 245-250.
Tanaka, M. & Tsuda, K. (2011). *Journal of Electron Microscopy* **60**, S245-S267.
Terauchi, M. & Tanaka, M. (1985). *Journal of Electron Microscopy* **34**, 128-135.
Venevtsev, Y. N., Politova, E. D. & Zhdanov, G. S. (1974). *Ferroelectrics* **8**, 489-490.

Vincent, R. & Midgley, P. A. (1994). *Ultramicroscopy* **53**, 271-282.
Woodward, D. I. & Reaney, I. M. (2005). *Acta Crystallographica Section B* **61**, 387-399.
Zuo, J. M. (2004). *Reports on Progress in Physics* **67**, 2053-2103.